\documentclass[11pt,a4paper]{article}

\usepackage{amssymb}
\usepackage[dvips]{graphicx}
\usepackage{bm}

\begin{document}

\title{\bf New form of the NSVZ relation at the two-loop level}

\author{
V.Yu.Shakhmanov, K.V.Stepanyantz\\
{\small{\em Moscow State University}}, {\small{\em Faculty of Physics, Department  of Theoretical Physics}}\\
{\small{\em 119991, Moscow, Russia}}}

\maketitle

\begin{abstract}
Recently the exact NSVZ $\beta$-function was rewritten in the form of a relation between the $\beta$-function and the anomalous dimensions of the quantum gauge superfield, of the Faddeev--Popov ghosts, and of the matter superfields. It was also suggested that this form of the NSVZ equation follows from an underlying equation relating two-point Green functions of the theory. Here we demonstrate that this relation is satisfied at the two-loop level for the non-Abelian ${\cal N}=1$ supersymmetric gauge theories in the case of using the simplest (BRST non-invariant) version of the higher covariant derivative regularization. Consequently, the integrals giving the two-loop $\beta$-function can be reduced to the one-loop integrals giving the anomalous dimensions of the quantum gauge superfield, of the Faddeev--Popov ghosts, and of the matter superfields.
\end{abstract}

\unitlength=1cm

\section{Introduction}
\hspace*{\parindent}

The exact NSVZ $\beta$-function \cite{Novikov:1983uc,Jones:1983ip,Novikov:1985rd,Shifman:1986zi} is the famous all-order relation between the $\beta$-function and the anomalous dimension of the chiral matter superfields (in the representation $R$ of the gauge group),\footnote{Here we do not so far specify, how the renormalization group functions are defined.}

\begin{equation}\label{NSVZ_Original_Equation}
\beta(\alpha,\lambda) = - \frac{\alpha^2\Big(3 C_2 - T(R) + C(R)_i{}^j (\gamma_\phi)_j{}^i(\alpha,\lambda)/r\Big)}{2\pi(1 - C_2\alpha/2\pi)}.
\end{equation}

\noindent
In our notation $\alpha=e^2/4\pi$ is the coupling constant and $\lambda^{ijk}$ are the Yukawa couplings. The group theory factors are defined by the equations $f^{ACD} f^{BCD} \equiv C_2 \delta^{AB}$, $\mbox{tr}\,(T^A T^B) \equiv T(R)\,\delta^{AB}$, and $(T^A)_i{}^k (T^A)_k{}^j \equiv C(R)_i{}^j$, where $(T^A)_i{}^j$ are the generators of the gauge group in the representation $R$. Generators $t^A$ of the fundamental representation in our notation satisfy the normalization condition $\mbox{tr}(t^A t^B) = \delta^{AB}/2$. The dimension of the gauge group is denoted by $r$.

For the pure ${\cal N}=1$ supersymmetric Yang--Mills (SYM) theory Eq. (\ref{NSVZ_Original_Equation}) gives the exact $\beta$-function in the form of the geometric series, and for theories with extended supersymmetry it produces all non-renormalization theorems \cite{Shifman:1999mv,Buchbinder:2014wra,Buchbinder:2015eva} which have been originally proposed in \cite{Grisaru:1982zh,Howe:1983sr} for ${\cal N}=2$ theories (see also Ref. \cite{Buchbinder:1997ib}) and in \cite{Grisaru:1982zh,Howe:1983sr,Mandelstam:1982cb,Brink:1982pd} for ${\cal N}=4$ SYM. The NSVZ-like relations also exist for the Adler $D$-function \cite{Adler:1974gd} in ${\cal N}=1$ SQCD \cite{Shifman:2014cya,Shifman:2015doa} and for theories with softly broken supersymmetry \cite{Hisano:1997ua,Jack:1997pa,Avdeev:1997vx}.

Although a lot of general arguments (see, e.g., \cite{Shifman:1999mv,ArkaniHamed:1997mj,Kraus:2002nu}) lead to Eq. (\ref{NSVZ_Original_Equation}), its direct derivation by methods of perturbation theory is a complicated and so far unsolved problem. Starting from the three-loop approximation the NSVZ relation is not valid in the $\overline{\mbox{DR}}$-scheme \cite{Jack:1996vg,Jack:1996cn,Jack:1998uj,Harlander:2006xq,Mihaila:2013wma} due to the scheme dependence \cite{Kutasov:2004xu,Kataev:2014gxa}. However, by the help of a specially tuned finite renormalization in a given loop it is possible to construct a scheme in which Eq. (\ref{NSVZ_Original_Equation}) is valid. Already in the three-loop approximation this fact is highly non-trivial \cite{Jack:1996vg}, because the NSVZ relation leads to some scheme independent consequences \cite{Kataev:2014gxa,Kataev:2013csa}. In the Abelian case the NSVZ scheme can be easily constructed for ${\cal N}=1$ theories regularized by the higher covariant derivative regularization \cite{Slavnov:1971aw,Slavnov:1972sq,Slavnov:1977zf} in a supersymmetric version \cite{Krivoshchekov:1978xg,West:1985jx}. This scheme is obtained (in all orders of perturbation theory) if only powers of $\ln\Lambda/\mu$ (where $\Lambda$ is the dimensionful parameter of the regularized theory and $\mu$ is a normalization point) are included in the renormalization constants \cite{Kataev:2013eta}. This prescription looks very similar to the minimal subtractions, so that it is possible to write

\begin{equation}\label{HDMS}
\mbox{NSVZ} = \mbox{HD}+\mbox{MSL},
\end{equation}

\noindent
where MSL means Minimal Subtractions of Logarithms. The NSVZ-like schemes for the Adler $D$-function in ${\cal N}=1$ SQCD \cite{Kataev:2017qvk} and for the renormalization of photino mass in ${\cal N}=1$ SQED \cite{Nartsev:2016mvn} are obtained by the same way.

The possibility of formulating this simple prescription giving the NSVZ scheme with the higher derivative regularization follows from an interesting feature of quantum corrections which was first noted in \cite{Soloshenko:2003nc,Smilga:2004zr}. Namely, the momentum integrals giving the $\beta$-function (defined in terms of the bare couplings) can be presented in the form of integrals of total derivatives \cite{Soloshenko:2003nc} and even double total derivatives \cite{Smilga:2004zr}. In ${\cal N}=1$ SQED this allows obtaining the NSVZ relation for the renormalization group (RG) functions defined in terms of the bare coupling constant by taking the integral over the momentum of a matter loop. The all-loop derivation of the NSVZ equation by this method for ${\cal N}=1$ SQED was made in Refs. \cite{Stepanyantz:2011jy,Stepanyantz:2014ima}  and has been verified by an explicit three-loop calculation in \cite{Kazantsev:2014yna}. Similar all-loop derivations of the NSVZ-like relations were done for the Adler $D$-function in ${\cal N}=1$ SQCD \cite{Shifman:2014cya,Shifman:2015doa} and for the renormalization of the photino mass in softly broken ${\cal N}=1$ SQED \cite{Nartsev:2016nym}. Note that the higher derivative regularization is a very essential ingredient of this construction, because for factorizing loop integrals into integrals of double total derivatives one should take the limit of the vanishing external momentum. For theories regularized by the dimensional reduction \cite{Siegel:1979wq,Siegel:1980qs} one should either introduce auxiliary masses in propagators in this limit \cite{Vladimirov:1979zm}, or make calculations for the non-vanishing external momentum. In the latter case the factorization into double total derivatives does not take place, although some similar constructions can be found \cite{Aleshin:2015qqc,Aleshin:2016rrr}.

The derivation of the NSVZ relation in all cases mentioned above has a simple graphical interpretation \cite{Smilga:2004zr} (see also \cite{Pimenov:2006cu}). Let us draw a supergraph without external lines. By attaching two external gauge legs in all possible ways we obtain a group of diagrams contributing to the $\beta$-function. The NSVZ equation relates it to the contribution to the anomalous dimension given by superdiagrams which are obtained from the original graph by cutting matter lines in all possible ways.

In the non-Abelian case explicit calculations in lowest loops with the higher covariant derivative regularization (see, e.g., \cite{Pimenov:2009hv,Steklov,Stepanyantz:2011bz,Stepanyantz:2012zz,Aleshin:2016yvj,Shakhmanov:2017soc}) also reveal factorization of integrals giving the $\beta$-function into integrals of double total derivatives. However, the above described graphical interpretation of Eq. (\ref{NSVZ_Original_Equation}) is evidently unapplicable in this case. Really, the $\beta$-function is obtained from diagrams with two external legs of the background superfield, while by cutting various propagators we obtain diagrams with external lines of the quantum gauge superfield, of the Faddeev-Popov ghosts, and of the matter superfields. Moreover, Eq. (\ref{NSVZ_Original_Equation}) contains the coupling constant dependent denominator which hinders comparing different contributions. All this difficulties were overcome in \cite{Stepanyantz:2016gtk}, where it was proved that the three-point gauge-ghost vertices are finite in all loops. (It is important that the gauge leg in these vertices corresponds to the {\it quantum} gauge superfield.) Using this non-renormalization theorem one can rewrite the NSVZ relation (\ref{NSVZ_Original_Equation}) (for the RG functions defined in terms of the bare couplings) in the equivalent form

\begin{equation}\label{NSVZ_New_Form}
\frac{\beta(\alpha_0,\lambda_0)}{\alpha_0^2} = - \frac{1}{2\pi}\Big(3 C_2 - T(R) - 2C_2 \gamma_c(\alpha_0,\lambda_0) - 2C_2 \gamma_V(\alpha_0,\lambda_0) + C(R)_i{}^j (\gamma_\phi)_j{}^i(\alpha_0,\lambda_0)/r\Big).
\end{equation}

\noindent
This equation relates the $\beta$-function to the anomalous dimensions $\gamma_c$, $\gamma_V$, and $(\gamma_\phi)_i{}^j$ of the Faddeev--Popov ghosts, of the quantum gauge superfield, and of the matter superfields, respectively. Moreover, it does not contain the coupling dependent denominator. That is why Eq. (\ref{NSVZ_New_Form}) admits exactly the same qualitative graphical interpretation as in the Abelian case.

Eq. (\ref{NSVZ_New_Form}) allows suggesting that, similar to the Abelian case, with higher covariant derivative regularization the Green functions (defined as in Ref.
\cite{Stepanyantz:2016gtk}, see Eqs. (\ref{Definition_Of_D}) and (\ref{Two_Point_Functions}) below) satisfy the equation

\begin{eqnarray}\label{NSVZ_For_Green_Functions}
&& \frac{d}{d\ln\Lambda}\Big(d^{-1} - \alpha_0^{-1}\Big)\Big|_{\alpha,\lambda=\mbox{\scriptsize const};\ p\to 0}
= - \frac{3 C_2 - T(R)}{2\pi}\nonumber\\
&&\qquad\quad - \frac{1}{2\pi}\frac{d}{d\ln\Lambda}\Big(- 2C_2 \ln G_c - C_2 \ln G_V + C(R)_i{}^j \ln (G_\phi)_j{}^i/r\Big)\Big|_{\alpha,\lambda=\mbox{\scriptsize const}; q\to 0}.\qquad
\end{eqnarray}

\noindent
If this equation is valid in all loops, than Eq. (\ref{NSVZ_New_Form}) is also valid, and Eq. (\ref{HDMS}) giving the NSVZ scheme takes place in the non-Abelian case. At the three-loop level this has been verified by an explicit calculation in Ref. \cite{Shakhmanov:2017soc} for terms quartic in the Yukawa couplings. However, this check works only for the matter part of Eq. (\ref{NSVZ_For_Green_Functions}), while it is much more interesting to compare parts coming from the contributions of the quantum gauge superfield and ghosts. This is much more difficult from technical point of view, because the higher covariant derivative term leads to new vertices of the complicated structure. That is why, at present, the complete calculation of the considered Green functions with the BRST-invariant version of the higher covariant derivative regularization was made only in the one-loop approximation \cite{Aleshin:2016yvj,Kazantsev:2017fdc}. Nevertheless, the calculations can be considerably simplified by the help of the BRST non-invariant versions of the higher derivative regularization, see Refs. \cite{Pimenov:2009hv,Steklov,Stepanyantz:2011bz,Stepanyantz:2012zz}. Such a non-invariant regularization should be supplemented by a special renormalization prescription which restores the Slavnov--Taylor identities \cite{Taylor:1971ff,Slavnov:1972fg} in each order of perturbation theory. Such prescriptions were constructed, e.g., in \cite{Slavnov:2001pu,Slavnov:2002ir} for non-supersymmetric gauge theories and in \cite{Slavnov:2002kg,Slavnov:2003cx} for the supersymmetric case.

In this paper we verify the relation (\ref{NSVZ_For_Green_Functions}) for the two-loop two-point Green function of the background gauge superfield and the one-loop two-point Green functions of the quantum gauge superfield, ghosts, and matter superfields. We take the two-loop result obtained with the BRST non-invariant version of the higher derivative regularization from Ref. \cite{Steklov} and compare it with the one-loop two-point Green functions, which are obtained here with the same regularization supplemented by the renormalization prescription proposed in \cite{Slavnov:2003cx}. This regularization is described in Sect. \ref{Section_Theory}. In Sect. \ref{Section_Calculation} we demonstrate the validity of Eq. (\ref{NSVZ_For_Green_Functions}) as an equality between the loop integrals.

\section{${\cal N}=1$ supersymmetric gauge theories regularized by higher derivatives}
\hspace*{\parindent}\label{Section_Theory}

We consider the ${\cal N}=1$ SYM theory interacting with the massless chiral superfields $\phi_i$ in a certain representation $R$ of the gauge group $G$,

\begin{eqnarray}
&& S = \frac{1}{2e_0^2} \mbox{Re}\, \mbox{tr}\int d^4x\,d^2\theta\, W^a W_a + \frac{1}{4} \int d^4x\, d^4\theta\, \phi^{*i} (e^{2V})_i{}^j \phi_j
\nonumber\\
&&\qquad\qquad\qquad\qquad\qquad\qquad\qquad\qquad\qquad
+ \Big(\frac{1}{6} \lambda_0^{ijk} \int d^4x\, d^2\theta\, \phi_i \phi_j \phi_k + \mbox{c.c.} \Big).\qquad
\end{eqnarray}

\noindent
The subscript $0$ denotes bare couplings, namely, the gauge coupling constant $e_0$ and the Yukawa couplings $\lambda_0^{ijk}$. Certainly, we assume that this theory is gauge invariant, so that the Yukawa couplings satisfy the equation

\begin{equation}
\lambda_0^{mjk} (T^A)_m{}^i + \lambda_0^{imk} (T^A)_m{}^j + \lambda_0^{ijm} (T^A)_m{}^k = 0.
\end{equation}

In this paper we will use almost the same version of the higher derivative regularization as in Ref. \cite{Steklov} (see also \cite{Stepanyantz:2011bz,Stepanyantz:2012zz}), because in this case we know all integrals defining the two-loop $\beta$-function. To construct this regularization, first, we make the background-quantum splitting by the help of the substitution $e^{2V} \to e^{\bm{\Omega}^+} e^{2V} e^{\bm{\Omega}}$. Then the background superfield $\bm{V}$ is given by the equation $e^{2\bm{V}} = e^{\bm{\Omega}^+} e^{\bm{\Omega}}$. It is convenient to choose the gauge fixing term which does not break the background gauge invariance, namely,\footnote{The gauge term (\ref{Gauge_Fixing_Term}) corresponds to the Feynman gauge $\xi=1$.}

\begin{equation}\label{Gauge_Fixing_Term}
S_{\mbox{\scriptsize gf}} = - \frac{1}{32 e^2} \mbox{tr} \int d^4x\, d^4\theta\, V\big(\bm{\nabla}^2 \bm{\bar\nabla}^2 + \bm{\bar\nabla}^2 \bm{\nabla}^2 \big)V,
\end{equation}

\noindent
where the background supersymmetric covariant derivatives are written as

\begin{equation}
\bm{\nabla}_a = e^{-\bm{\Omega}^+} D_a e^{\bm{\Omega}^+};\qquad \bm{\bar\nabla}_{\dot a} = e^{\bm{\Omega}} D_{\dot a} e^{-\bm{\Omega}}.
\end{equation}

\noindent
Certainly, it is also necessary to introduce anticommuting Faddeev--Popov and Nielsen--Kallosh ghost superfields with the actions

\begin{eqnarray}
&& S_{\mbox{\scriptsize FP}} = \frac{1}{e_0^2}\mbox{tr} \int d^4x\, d^4\theta\,\left(e^{\bm{\Omega}} \bar c e^{-\bm{\Omega}} + e^{-\bm{\Omega}^+} \bar c^+ e^{\bm{\Omega}^+}\right)
\nonumber\\
&&\qquad\qquad\qquad
\times \Big\{\Big(\frac{V}{1-e^{2V}}\Big)_{Adj} \left(e^{-\bm{\Omega}^+} c^+ e^{\bm{\Omega}^+}\right)
+ \Big(\frac{V}{1-e^{-2V}}\Big)_{Adj} \left(e^{\bm{\Omega}} c e^{-\bm{\Omega}}\right)\Big\};\qquad\\
&& S_{\mbox{\scriptsize NK}} = \frac{1}{2e_0^2} \mbox{tr} \int d^4x\,d^4\theta\, b^+ e^{2\bm{V}} b e^{-2\bm{V}},
\end{eqnarray}

\noindent
respectively.

In this paper we use a version of the higher covariant derivative regularization which is obtained by adding

\begin{equation}\label{Higher_Derivative_Term}
S_\Lambda = \frac{1}{2 e_0^2}  \mbox{Re}\, \mbox{tr}\int d^4x\,d^4\theta\, V \frac{(\bm{\nabla}_\mu^2)^{n+1}}{\Lambda^{2n}} V
+ \frac{1}{4} \int d^4x\, d^4\theta\, \phi^{*i} \Big(e^{\bm{\Omega}^+}\frac{(\bm{\nabla}_\mu^2)^{m}}{\Lambda^{2m}} e^{\bm{\Omega}}\Big)_i{}^j \phi_j
\end{equation}

\noindent
to the action, where the background covariant derivative $\bm{\nabla}_\mu$ is defined by the equation

\begin{equation}
\{\bm{\nabla}_a, \bm{\bar \nabla}_{\dot b} \} = 2i (\gamma^\mu)_{a \dot b} \bm{\nabla}_\mu.
\end{equation}

\noindent
By adding the term (\ref{Higher_Derivative_Term}) we regularize all divergences except for the ones in the one-loop approximation, for which one should use the Pauli--Villars method \cite{Slavnov:1977zf}. To cancel the one-loop divergences, it is possible to insert the Pauli--Villars determinants for the matter superfields and ghosts,\footnote{In the Feynman gauge one-loop diagrams with a loop of the quantum gauge superfield are finite.}

\begin{equation}\label{Pauli_Villrs_Determinants}
\Big(\int D\Phi\, \exp\left(iS_\Phi\right)\Big)^{-1} \int DB \exp\left(iS_B\right) \prod\limits_{I=1}^{K} \Big(\int D\bar C_I\, DC_I\,\exp\left(i S_{C,I}\right)\Big)^{c_I},
\end{equation}

\noindent
into the generating functional. Here the (commuting) Pauli--Villars superfields $\Phi_i$ with the action

\begin{equation}
S_\Phi = \frac{1}{4} \int d^4x\, d^4\theta\,\Phi^{*i} \Big(e^{\bm{\Omega}^+}\frac{(\bm{\nabla}_\mu^2)^{m}}{\Lambda^{2m}} e^{\bm{\Omega}}
+ e^{\bm{\Omega}^+} e^{2V} e^{\bm{\Omega}}\Big)_i{}^j \Phi_j + \Big(\frac{1}{4} M^{ij}\int d^4x\, d^2\theta\, \Phi_i\Phi_j + \mbox{c.c.}\Big)
\end{equation}

\noindent
lie in the representation $R$. Also we assume existence of the invariant mass term, such that $M^{ij} M^*_{jk} = \delta_k^i M^2$ with $M=a_\phi\Lambda$, where $a_\phi$ is a
constant. The chiral superfields $\bar C_I$ and $C_I$ with $I=1,\ldots,K$ are anticommuting and lie in the adjoint representation. The commuting chiral superfield $B$ also lies in the adjoint representation. The actions for these superfields have the form

\begin{eqnarray}
&& S_{C,I} = \frac{1}{e_0^2}\mbox{tr} \int d^4x\, d^4\theta\,\left(e^{\bm{\Omega}} \bar C_I e^{-\bm{\Omega}} + e^{-\bm{\Omega}^+} \bar C_I^+ e^{\bm{\Omega}^+}\right)
\Big\{\Big(\frac{V}{1-e^{2V}}\Big)_{Adj} \left(e^{-\bm{\Omega}^+} C_I^+ e^{\bm{\Omega}^+}\right)
\nonumber\\
&&\qquad\qquad\qquad\qquad
+ \Big(\frac{V}{1-e^{-2V}}\Big)_{Adj} \left(e^{\bm{\Omega}} C_I e^{-\bm{\Omega}}\right)\Big\}
+ \Big(\frac{m_{C,I}}{e_0^2}\mbox{tr}\int d^4x\, d^2\theta\, \bar C_I C_I + \mbox{c.c.}\Big);\qquad\\
&& S_B = \frac{1}{2e_0^2} \mbox{tr} \int d^4x\,d^4\theta\, B^+ e^{2\bm{V}} B e^{-2\bm{V}} + \Big(\frac{m_B}{2e_0^2}\mbox{tr} \int d^4x\, d^2\theta\, B^2 + \mbox{c.c.} \Big).
\end{eqnarray}

\noindent
The masses should be proportional to the parameter $\Lambda$, $m_B = a_B\Lambda$, $m_{C,I} = a_{C,I} \Lambda$, where the coefficients $a_B$ and $a_{C,I}$ are some constants independent of the couplings. To cancel the one-loop divergences, the coefficients $c_I$ in Eq. (\ref{Pauli_Villrs_Determinants}) should satisfy the conditions

\begin{equation}\label{Conditions_For_C_I}
1 + \sum\limits_{I=1}^K c_I = 0;\qquad \sum\limits_{I=1}^K c_I m_{C,I}^2 = 0.
\end{equation}

Note that the higher derivative term (\ref{Higher_Derivative_Term}) does not break the background gauge invariance and leads to relatively simple calculations, because there are no new vertices containing the quantum gauge superfield. However, the BRST invariance (which is a remnant of the quantum gauge invariance after the gauge fixing procedure \cite{Becchi:1974md,Tyutin:1975qk}) is broken. The BRST-invariant version of the higher covariant derivative regularization can be also constructed, but the calculations made by the help of it are much more complicated, see, e.g. \cite{Aleshin:2016yvj}. At present, the $\beta$-function of the general renormalizable ${\cal N}=1$ supersymmetric gauge theory has been calculated with the BRST invariant version of the higher derivative regularization only in the one-loop approximation \cite{Aleshin:2016yvj}, although some contributions are known even in the three-loop approximation \cite{Shakhmanov:2017soc}. For the regularization obtained with the higher derivative term (\ref{Higher_Derivative_Term}), all integrals giving the two-loop $\beta$-function have been found in Refs. \cite{Steklov,Stepanyantz:2011bz,Stepanyantz:2012zz}. Certainly, the non-invariant regularization should be supplemented by a proper subtraction scheme which restores the Slavnov--Taylor identities. In this paper we will use a procedure similar to the one proposed in \cite{Slavnov:2003cx}.

\section{Relation between the Green functions}
\hspace*{\parindent}\label{Section_Calculation}

First, we remind the result for the two-loop two-point Green function of the background gauge superfield obtained in \cite{Steklov,Stepanyantz:2011bz}. Due to the unbroken
background gauge invariance the corresponding part of the effective action can be presented in the form

\begin{equation}\label{Definition_Of_D}
\Gamma^{(2)}_{\bm{V}} = - \frac{1}{8\pi} \mbox{tr} \int \frac{d^4p}{(2\pi)^4}\,d^4\theta\, \bm{V}(p,\theta) \partial^2\Pi_{1/2} \bm{V}(-p,\theta)\,
d^{-1}(\alpha_0,\lambda_0,\Lambda/p),
\end{equation}

\noindent
where $\partial^2\Pi_{1/2}$ is the supersymmetric transversal projection operator. In this equation the coefficient is chosen so that the function $d$ coincides with $\alpha_0 = e_0^2/4\pi$ in the tree approximation. In the two-loop approximation

\begin{eqnarray}\label{Beta_Two_Loop}
&& \frac{d}{d\ln\Lambda}\Big(d^{-1} - \alpha_0^{-1}\Big)\Big|_{\alpha,\lambda=\mbox{\scriptsize const};\ p\to 0} = C_2 \left(I_{\mbox{\scriptsize FP}}
+ I_{\mbox{\scriptsize NK}}\right) + T(R) I_0 + \alpha_0 (C_2)^2 I_1\nonumber\\
&& + \frac{\alpha_0}{r} \mbox{tr} \left(C(R)^2\right) I_2 + \alpha_0 C_2 T(R) I_3
+ C(R)_i{}^j \frac{\lambda_0^{imn}\lambda^*_{0jmn}}{4\pi r} I_4 + O(\alpha_0^2,\alpha_0\lambda_0^2,\lambda_0^4).\qquad
\end{eqnarray}

\noindent
Writing the (Euclidean) integrals $I_i$ in this equation we will use the notation

\begin{equation}
F_q \equiv 1 + \frac{q^{2m}}{\Lambda^{2m}};\qquad R_q \equiv 1 + \frac{q^{2n}}{\Lambda^{2n}}.
\end{equation}

\noindent
The one-loop contributions $I_{\mbox{\scriptsize FP}}$, $I_{\mbox{\scriptsize NK}}$, and $I_0$ come from diagrams with a loop of the ghost or matter superfields and
can be easily calculated,

\begin{eqnarray}
&& I_{\mbox{\scriptsize FP}} =  2\pi \sum\limits_{I=1}^K c_I \int \frac{d^4q}{(2\pi)^4} \frac{d}{d\ln\Lambda} \frac{\partial}{\partial q^\mu}
\frac{\partial}{\partial q_\mu} \Big[\frac{1}{q^2} \ln \Big(1+ \frac{m_{C,I}^2}{q^2}\Big)\Big] = -\frac{1}{\pi};\\
&& I_{\mbox{\scriptsize NK}} = - \pi \int \frac{d^4q}{(2\pi)^4} \frac{d}{d\ln\Lambda} \frac{\partial}{\partial q^\mu}
\frac{\partial}{\partial q_\mu} \Big[\frac{1}{q^2} \ln \Big(1+ \frac{m_B^2}{q^2}\Big)\Big] = -\frac{1}{2\pi};\quad\\
&& I_0 = \pi \int \frac{d^4q}{(2\pi)^4} \frac{d}{d\ln\Lambda} \frac{\partial}{\partial q^\mu}
\frac{\partial}{\partial q_\mu} \Big[\frac{1}{q^2} \ln \Big(1+ \frac{M^2}{q^2 F_q^2}\Big)\Big] = \frac{1}{2\pi}.
\end{eqnarray}

\noindent
They give the first term in the right hand side of Eq. (\ref{NSVZ_For_Green_Functions}).

The two-loop contributions to the $\beta$-function are given by the integrals

\begin{eqnarray}
&&\hspace*{-2mm} I_1 = -12\pi^2 \int \frac{d^4q}{(2\pi)^4} \frac{d^4k}{(2\pi)^4} \frac{d}{d\ln\Lambda} \frac{\partial}{\partial q^\mu}
\frac{\partial}{\partial q_\mu} \Big[\frac{1}{k^2 R_k q^2 R_q (q+k)^2 R_{q+k}}\Big];\\
&&\hspace*{-2mm} I_2 = 8\pi^2 \int \frac{d^4q}{(2\pi)^4} \frac{d^4k}{(2\pi)^4} \frac{d}{d\ln\Lambda} \frac{\partial}{\partial q^\mu}
\frac{\partial}{\partial q_\mu} \Big[\frac{1}{k^2 R_k q^2 F_q (q+k)^2 F_{q+k}}
\nonumber\\
&&\hspace*{-2mm}\qquad\qquad\qquad\qquad\qquad\qquad\qquad\qquad\quad
- \frac{F_q F_{q+k}}{k^2 R_k \left(q^2 F_q^2 + M^2\right)\left((q+k)^2 \smash{F_{q+k}^2}\vphantom{F_q^2} + M^2\right)}\Big];\qquad\\
&&\hspace*{-2mm} I_3 = 8\pi^2 \int \frac{d^4q}{(2\pi)^4} \frac{d^4k}{(2\pi)^4} \frac{d}{d\ln\Lambda} \frac{\partial}{\partial q^\mu}
\frac{\partial}{\partial k_\mu} \Big[\frac{1}{(k+q)^2 R_{k+q} q^2 F_q k^2 F_k}
\nonumber\\
&&\hspace*{-2mm}\qquad\qquad\qquad\qquad\qquad\qquad\qquad\qquad\quad
- \frac{F_q F_k}{(k+q)^2 R_{k+q} \left(q^2 F_q^2 + M^2\right)\left(k^2 F_k^2 + M^2\right)}\Big];\qquad\\
&&\hspace*{-2mm} I_4 = -8\pi^2 \int \frac{d^4q}{(2\pi)^4} \frac{d^4k}{(2\pi)^4} \frac{d}{d\ln\Lambda} \frac{\partial}{\partial q^\mu}
\frac{\partial}{\partial q_\mu} \Big[\frac{1}{k^2 F_k q^2 F_q (q+k)^2 F_{q+k}}\Big].
\end{eqnarray}

\noindent
All of them are integrals of double total derivatives. This allows taking one of the momentum integrals,

\begin{eqnarray}\label{I1}
&& I_1 = -6 \int \frac{d^4k}{(2\pi)^4} \frac{d}{d\ln\Lambda} \frac{1}{k^4 R_k^2};\\
\label{I2}
&& I_2 = 4 \int \frac{d^4k}{(2\pi)^4} \frac{d}{d\ln\Lambda} \frac{1}{k^4 F_k R_k};\\
\label{I3}
&& I_3 = 2 \int \frac{d^4k}{(2\pi)^4} \frac{d}{d\ln\Lambda} \Big[\frac{1}{k^4 F_k^2} - \frac{F_k^2}{\left(k^2 F_k^2 + M^2\right)^2}\Big];\\
\label{I4}
&& I_4 = -4 \int \frac{d^4k}{(2\pi)^4} \frac{d}{d\ln\Lambda} \frac{1}{k^4 F_k^2}.
\end{eqnarray}

\noindent
Certainly, these integrals can be easily calculated (see, e.g., Ref. \cite{Stepanyantz:2011bz}), but we will not do this, because here we would like to verify Eq.
(\ref{NSVZ_For_Green_Functions}) at the level of loop integrals. Using Eqs. (\ref{Beta_Two_Loop}) and (\ref{I1})--(\ref{I4}) the left hand side of Eq.
(\ref{NSVZ_For_Green_Functions}) can be written as

\begin{eqnarray}\label{Beta_Result}
&& \frac{d}{d\ln\Lambda}\Big(d^{-1} - \alpha_0^{-1}\Big)\Big|_{\alpha,\lambda=\mbox{\scriptsize const};\ p\to 0} = - \frac{3 C_2 - T(R)}{2\pi}
+  \int \frac{d^4k}{(2\pi)^4} \frac{1}{k^4} \frac{d}{d\ln\Lambda} \left[\vphantom{\frac{1}{2}}\right. - (C_2)^2 \frac{6\alpha}{R_k^2}\qquad\nonumber\\
&& + \mbox{tr} \left(C(R)^2\right) \frac{4\alpha}{r F_k R_k} + 2\alpha C_2 T(R)
\left(\vphantom{\frac{1}{2}}\smash{\frac{1}{F_k^2} - \frac{k^4 F_k^2}{\left(k^2 F_k^2 + M^2\right)^2}}\right)
- C(R)_i{}^j \frac{\lambda^{imn}\lambda^*_{jmn}}{\pi r F_k^2} \left.\vphantom{\frac{1}{2}}\right]\nonumber\\
&& + O(\alpha^2,\alpha\lambda^2,\lambda^4),\vphantom{\frac{1}{2}}
\end{eqnarray}

\noindent
where we take into account that $\alpha_0=\alpha + O(\alpha^2)$ and $\lambda_0^{ijk} = \lambda^{ijk} + O(\alpha\lambda,\lambda^3)$.

Let us compare this expression with the one-loop two-point Green functions of the quantum gauge superfield, of the Faddeev--Popov ghosts, and of the matter superfields.
The functions $G_V$, $G_c$, and $(G_\phi)_j{}^i$ entering Eq. (\ref{NSVZ_For_Green_Functions}) are related to the corresponding contributions to the effective action by the
equation

\begin{eqnarray}\label{Two_Point_Functions}
&& \Gamma^{(2)} - S_{\mbox{\scriptsize gf}}^{(2)} =
-\frac{1}{2e_0^2} \mbox{tr} \int \frac{d^4q}{(2\pi)^4}\, d^4\theta\, V(q,\theta) \partial^2\Pi_{1/2} V(-q,\theta)\, G_V(\alpha_0,\lambda_0,\Lambda/q)\nonumber\\
&& + \frac{1}{2e_0^2}\mbox{tr} \int \frac{d^4q}{(2\pi)^4}\, d^4\theta\, \Big(-\bar c(q,\theta) c^+(-q,\theta) + \bar c^+(q,\theta)
c(-q,\theta)\Big) G_c(\alpha_0,\lambda_0,\Lambda/q)\quad\nonumber\\
&& + \frac{1}{4} \int \frac{d^4q}{(2\pi)^4}\, d^4\theta\, \phi^{*j}(q,\theta) \phi_i(-q,\theta)\,
(G_\phi)_j{}^i(\alpha_0,\lambda_0,\Lambda/q) + \ldots,
\end{eqnarray}

\noindent
where the dots denote the other terms in the effective action quadratic in fields. According to this definition, in the tree approximation

\begin{equation}
G_V = 1 + O(\alpha_0);\qquad G_c = 1 + O(\alpha_0);\qquad (G_\phi)_i{}^j = \delta_i{}^j + O(\alpha_0,\lambda_0^2).
\end{equation}

Note that the two-point Green function of the quantum gauge superfield is transversal. This follows from the Slavnov--Taylor identities \cite{Taylor:1971ff,Slavnov:1972fg} and the Schwinger--Dyson equations for the Faddeev--Popov ghosts. However, the regularization used in this paper breaks the BRST invariance of the total action and, therefore, the Slavnov--Taylor identities are valid only after a special renormalization procedure made according to the prescription presented in \cite{Slavnov:2003cx}. In particular, the two-point Green function of the quantum gauge superfield will be transversal only after this procedure. Below we will demonstrate this by an explicit calculation.

Let us calculate the derivatives of the functions $(G_\phi)_j{}^i$, $G_c$, and $G_V$ with respect to $\ln\Lambda$ in the limit of the vanishing external momentum with the considered version of the higher derivative regularization.\footnote{Earlier such a calculation has been made only with the BRST invariant version of the higher covariant derivative regularization in \cite{Aleshin:2016yvj}, but here it is important that various Green functions will be obtained with the same regularization method.}

\begin{figure}[h]
\begin{picture}(0,2)
\put(2.1,0.2){\includegraphics[scale=0.42]{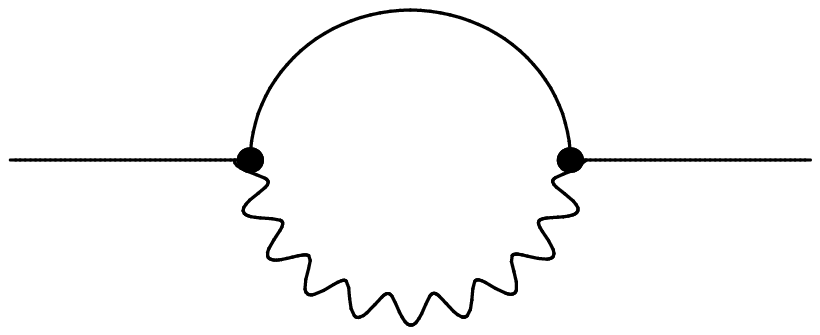}}
\put(6,-0.3){\includegraphics[scale=0.40]{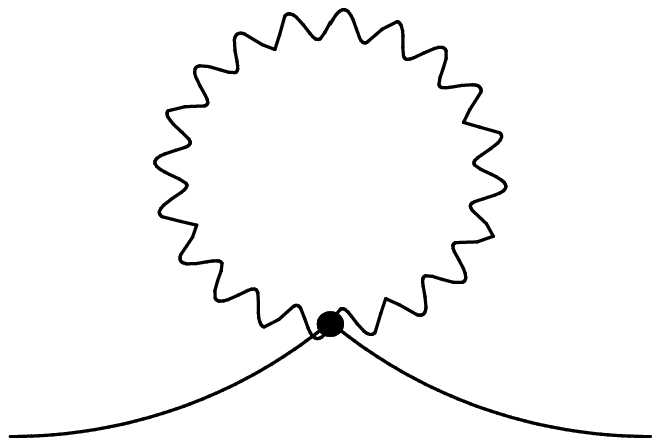}}
\put(10.0,0.2){\includegraphics[scale=0.42]{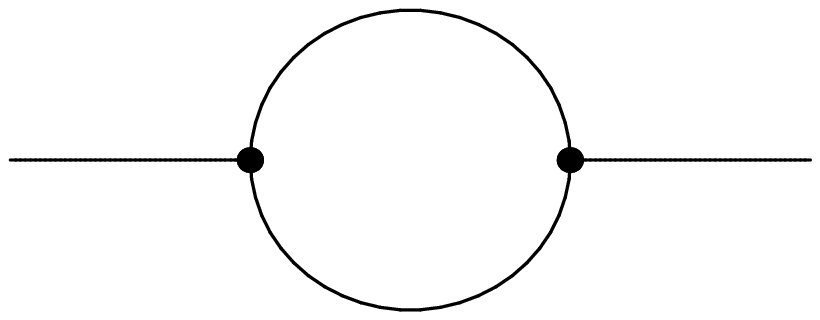}}
\end{picture}
\caption{Superdiagrams giving the two-point Green function of the matter superfields in the one-loop approximation.}\label{Figure_Phi}
\end{figure}

The one-loop two-point Green function of the matter superfields is obtained by calculating the diagrams presented in Fig. \ref{Figure_Phi}. This calculation gives

\begin{eqnarray}
&& \frac{d}{d\ln\Lambda} \ln (G_\phi)_j{}^i\Big|_{\alpha,\lambda=\mbox{\scriptsize const};\ q\to 0}
= \int \frac{d^4k}{(2\pi)^4} \left(-C(R)_j{}^i \frac{8\pi\alpha_0}{k^4 F_k R_k} + \lambda_0^{imn}\lambda^*_{0jmn} \frac{2}{k^4 F_k^2}\right)\qquad\nonumber\\
&& + O(\alpha_0^2,\alpha_0\lambda_0^2,\lambda_0^4).\vphantom{\frac{1}{2}}
\end{eqnarray}

\begin{figure}[h]
\begin{picture}(0,2)
\put(4.3,0.2){\includegraphics[scale=0.42]{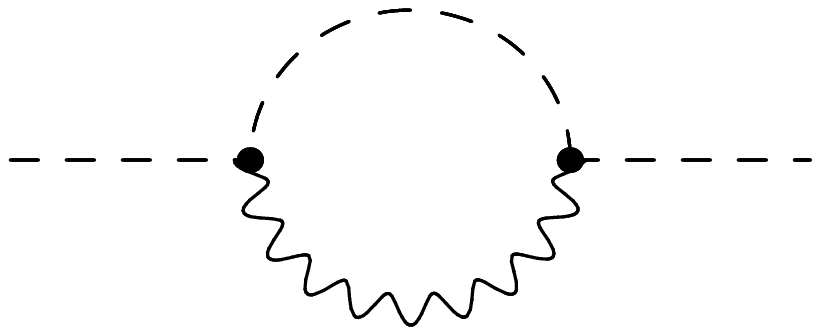}}
\put(8.7,0.0){\includegraphics[scale=0.41]{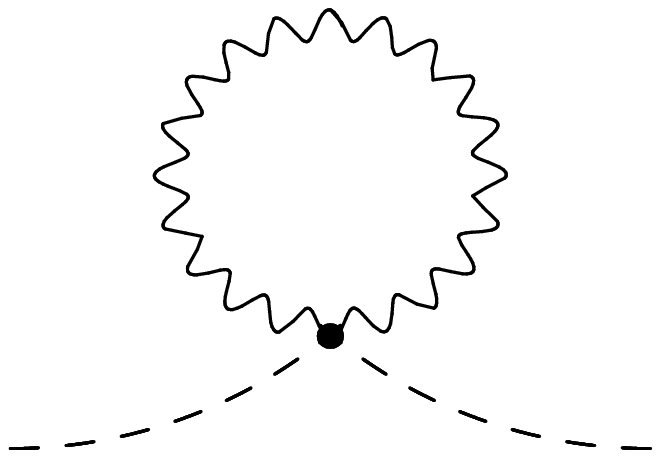}}
\end{picture}
\caption{Superdiagrams contributing to the one-loop two-point Green function of the Faddeev--Popov ghosts.}\label{Figure_FP_Ghosts}
\end{figure}

The superdiagrams contributing to the one-loop two-point Green function of the Faddeev--Popov ghosts are presented in Fig. \ref{Figure_FP_Ghosts}. In the Feynman
gauge $\xi=1$ (which is used in this paper following Refs. \cite{Steklov,Stepanyantz:2011bz}) their sum in the limit of the vanishing external momentum is 0,

\begin{equation}
\frac{d}{d\ln\Lambda} \ln G_c\Big|_{\alpha,\lambda=\mbox{\scriptsize const};\ q\to 0} = O(\alpha_0^2).
\end{equation}

\begin{figure}[h]
\begin{picture}(0,4)
\put(2.1,2.4){\includegraphics[scale=0.4]{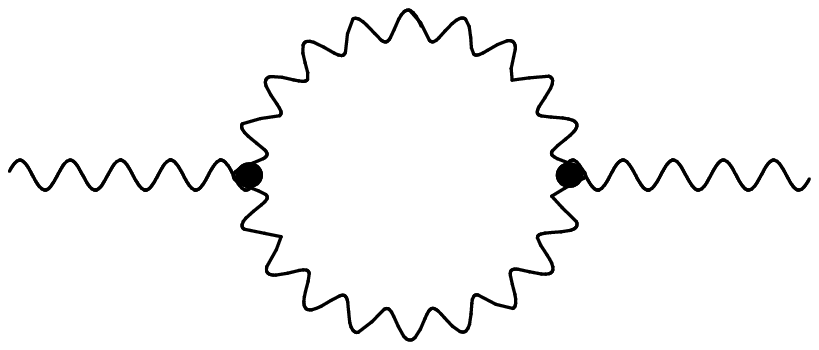}}
\put(2.5,0){\includegraphics[scale=0.4]{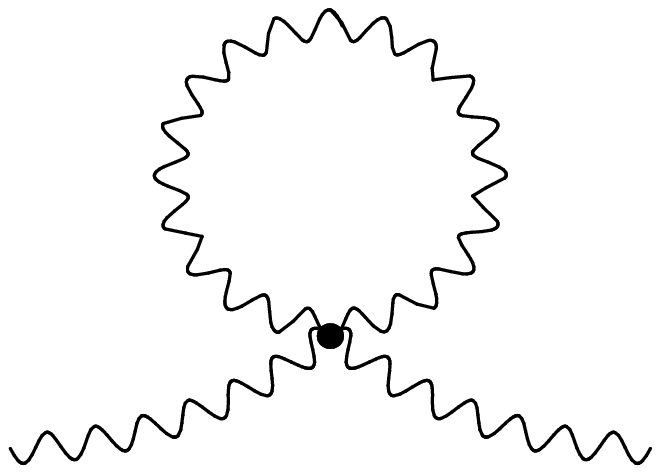}}
\put(6.4,2.43){\includegraphics[scale=0.4]{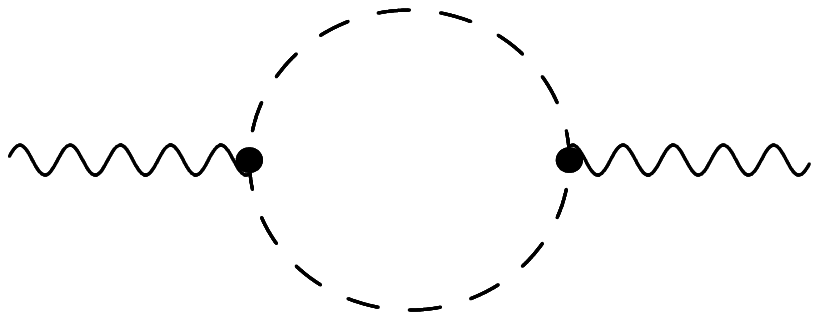}}
\put(6.7,-0.01){\includegraphics[scale=0.4]{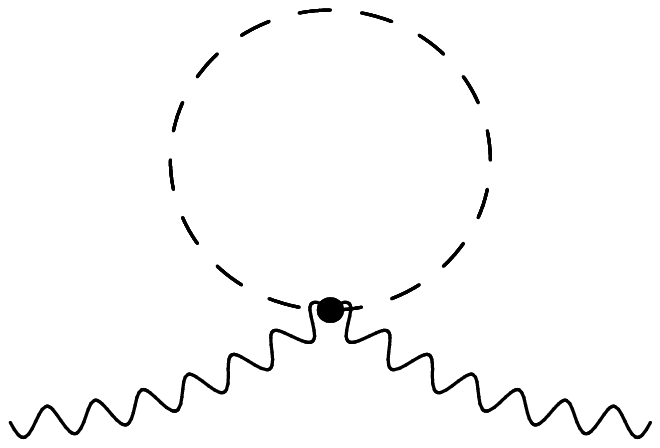}}
\put(10.7,2.4){\includegraphics[scale=0.4]{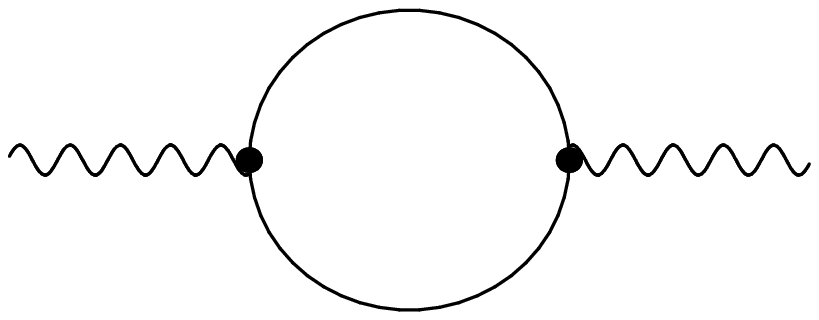}}
\put(11.1,-0.02){\includegraphics[scale=0.4]{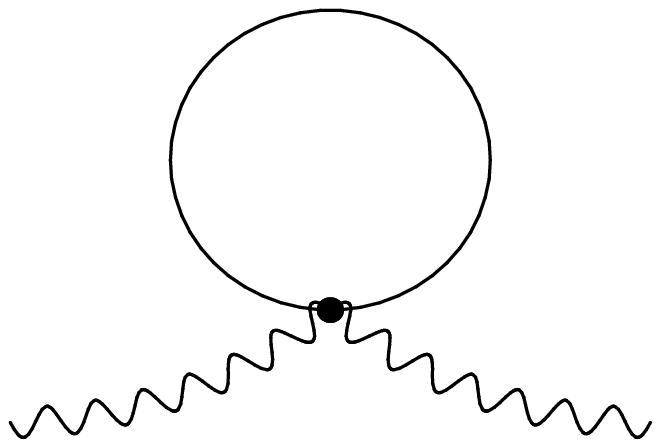}}
\end{picture}
\caption{Superdiagrams giving the two-point Green function of the quantum gauge superfield $V$ in the one-loop approximation.}
\label{Figure_QuantumV}
\end{figure}

The two-point Green function of the quantum gauge superfield in the one-loop approximation is contributed to by the superdiagrams presented in Fig. \ref{Figure_QuantumV}. The result for them is written as

\begin{eqnarray}\label{Two_Point_V}
&& \Gamma^{(2)}_V -S^{(2)}_{\mbox{\scriptsize gf}}
= -\frac{1}{2e_0^2} \mbox{tr} \int \frac{d^4q}{(2\pi)^4}\, d^4\theta\,\Big(V(q,\theta) \partial^2\Pi_{1/2} V(-q,\theta)\, G_V(\alpha_0,\lambda_0,\Lambda/q)\nonumber\\
&&\qquad\qquad\qquad\qquad\qquad\qquad\qquad\qquad\qquad\qquad\quad + V(q,\theta) V(-q,\theta)\,\widetilde G_V(\alpha_0,\lambda_0,\Lambda/q) \Big).\qquad\quad
\end{eqnarray}

\noindent
Unlike Eq. (\ref{Two_Point_Functions}), the Green function obtained from this equation is not transversal. This occurs, because the regularization breaks the BRST
invariance. Consequently, the Slavnov--Taylor identities are also broken and the two-point Green function of the quantum gauge superfield is no longer
transversal.\footnote{Note that the transversality of the two-point Green function of the background gauge superfield is ensured by the unbroken background gauge invariance.} Explicitly calculating the diagrams presented in Fig. \ref{Figure_QuantumV} we obtain that the non-invariant part of the two-point function (\ref{Two_Point_V})
in the limit of the vanishing external momentum is written as

\begin{eqnarray}
&&\hspace*{-5mm} \widetilde G_V\Big|_{q\to 0} = -8\pi\alpha_0 \int \frac{d^4k}{(2\pi)^4} \Big[ \frac{1}{6} C_2\Big(\frac{1}{k^2}+\sum\limits_{I=1}^K c_I \frac{1}{k^2+m_{C,I}^2}
-\frac{1}{k^2 R_k}\Big) + T(R) \frac{(F_k-1) M^2}{k^2 F_k^2 (k^2 F_k^2 + M^2)} \Big]\nonumber\\
&&\hspace*{-5mm} + O(\alpha_0^2,\alpha_0\lambda_0^2) \equiv e_0^2 \Lambda^2 g_V + O(\alpha_0^2,\alpha_0\lambda_0^2).
\end{eqnarray}

\noindent
For $m\ge 1$ and $n>1$ this integral is convergent due to the conditions (\ref{Conditions_For_C_I}), so that (for finite $\Lambda$) $g_V$ is a finite numerical constant.

According to Ref. \cite{Slavnov:2001pu,Slavnov:2002ir}, starting from the result obtained with a non-invariant regularization one should construct Green functions satisfying the Slavnov--Taylor identities by the help of a special algorithm. In the supersymmetric case this algorithm is described in \cite{Slavnov:2002kg} for the Abelian theories, and in \cite{Slavnov:2003cx} for the non-Abelian ones. The first step of this algorithm is to remove non-invariant terms proportional to $V(q,\theta) V(-q,\theta)$ from the two-point Green function of the quantum gauge superfield and keep only invariant terms proportional to $V(q,\theta) \partial^2\Pi_{1/2} V(-q,\theta)$. Consequently, the Slavnov--Taylor identity (in this case, the transversality of the considered Green function) is satisfied by the renormalized Green function. It is easy to demonstrate that the above procedure is equivalent to adding the non-invariant counterterm

\begin{equation}
\Delta S = \frac{1}{2} g_V \Lambda^2\, \mbox{tr} \int d^4x\, d^4\theta\, V^2,
\end{equation}

\noindent
which appears due to the use of the non-invariant regularization.\footnote{Certainly, there is also a non-invariant counterterm containing $V\partial^2 V$, but to find it, it is necessary to calculate the function $\widetilde G_V$ for non-vanishing values of $q$. Such calculations are much more complicated from the technical point of view, see, e.g., \cite{Kazantsev:2017fdc}.}

The invariant part of the function (\ref{Two_Point_V}) is logarithmically divergent in the limit $\Lambda\to \infty$. After calculating the diagrams presented in Fig. \ref{Figure_QuantumV} we have obtained

\begin{eqnarray}
&& \frac{d}{d\ln\Lambda} \ln G_V\Big|_{\alpha,\lambda=\mbox{\scriptsize const};\ q\to 0} = \pi \alpha_0 \int \frac{d^4k}{(2\pi)^4} \frac{d}{d\ln\Lambda}
\left[\vphantom{\frac{1}{2}}\right. - \frac{12 C_2}{k^4 R_k^2}\nonumber\\
&&\qquad\qquad\qquad\qquad\qquad\qquad + T(R)\left(\frac{4}{k^4 F_k^2}\right.
\left. - \left.\smash{\frac{4 F_k^2}{\left(k^2 F_k^2 + M^2\right)^2}}\vphantom{\frac{1}{2}}\right)\right] + O(\alpha_0^2,\alpha_0\lambda_0^2).\qquad
\end{eqnarray}

Thus, the terms containing various $\ln G$ in the right hand side of Eq. (\ref{NSVZ_For_Green_Functions}) can be written as

\begin{eqnarray}\label{Gamma_Result}
&& - \frac{1}{2\pi}\frac{d}{d\ln\Lambda}\Big(- 2C_2 \ln G_c - C_2 \ln G_V + C(R)_i{}^j \ln (G_\phi)_j{}^i/r\Big)
\Big|_{\alpha,\lambda=\mbox{\scriptsize const}; q\to 0}\nonumber\\
&& = \int \frac{d^4k}{(2\pi)^4} \frac{1}{k^4} \frac{d}{d\ln\Lambda} \left[\vphantom{\frac{1}{2}}\right. - (C_2)^2 \frac{6\alpha}{R_k^2}
+ 2\alpha C_2 T(R) \left(\vphantom{\frac{1}{2}}\smash{\frac{1}{F_k^2} - \frac{k^4 F_k^2}{\left(k^2 F_k^2 + M^2\right)^2}}\right)\nonumber\\
&&\qquad\qquad\qquad\qquad\quad\ \ + \mbox{tr} \left(C(R)^2\right) \frac{4\alpha}{r F_k R_k} - C(R)_i{}^j \frac{\lambda^{imn}\lambda^*_{jmn}}{\pi r F_k^2} \left.\vphantom{\frac{1}{2}}\right]
+ O(\alpha^2,\alpha\lambda^2,\lambda^4).\qquad\quad
\end{eqnarray}

\noindent
Comparing this expression with Eq. (\ref{Beta_Result}) we see that the NSVZ-like equation (\ref{NSVZ_For_Green_Functions}) relating various two-point Green functions is
really valid in the considered approximation with the considered version of the higher derivative regularization. (Note that the equality takes place for loop integrals
in the case of arbitrary values of the parameters $m$ and $n>1$.) This confirms the assumption made in Ref. \cite{Stepanyantz:2016gtk} that Eq. (\ref{NSVZ_For_Green_Functions}) is valid in all orders and can be used for deriving the NSVZ relation and constructing the NSVZ scheme in the non-Abelian case.

\section{Conclusion}
\hspace*{\parindent}

In this paper we have verified Eq. (\ref{NSVZ_For_Green_Functions}) for the general ${\cal N}=1$ supersymmetric gauge theory by comparing the two-loop two-point Green function of the background gauge superfield with the one-loop two-point Green functions of the quantum gauge superfield, of the Faddeev--Popov ghosts, and of the matter superfields. To make this check we use the BRST non-invariant version of the higher derivative regularization supplemented by a subtraction scheme which restore the Slavnov--Taylor identities. This regularization was chosen, because in this case all integrals defining the two-loop running of the coupling constant have been calculated earlier. After calculating one-loop two-point Green functions of the quantum fields listed above, we have checked that Eq. (\ref{NSVZ_For_Green_Functions}) is really satisfied. This confirms the proposal made in Ref. \cite{Stepanyantz:2016gtk} that this equation is valid in all orders and can be used as a starting point for deriving the NSVZ relation (\ref{NSVZ_Original_Equation}) by direct summation of supergraphs. Nevertheless, it is highly desirable to verify Eq. (\ref{NSVZ_For_Green_Functions}) in the case of using the BRST invariant version of higher derivative regularization. This problem is more difficult from the technical point of view, because of new higher derivative vertices. These new vertices essentially complicate the calculation even of the two-loop $\beta$-function. However, we hope that in future it would be possible to make such a check of Eq. (\ref{NSVZ_For_Green_Functions}).

\section*{Acknowledgements}
\hspace*{\parindent}

We are very grateful to A.L.Kataev and A.E.Kazantsev for valuable discussions. Also we would like to express our thanks to A.L.Kataev for the idea to use the abbreviation MSL.

\end{document}